\documentclass{article}

\usepackage[final]{neurips_2021}

\usepackage[utf8]{inputenc} %
\usepackage[T1]{fontenc}    %
\usepackage{url}            %
\usepackage{booktabs}       %
\usepackage{amsfonts}       %
\usepackage{nicefrac}       %
\usepackage{microtype}      %
\usepackage{xcolor}         %

\usepackage{array}
\usepackage{graphicx}
\usepackage{clrscode}
\usepackage{enumitem}
\usepackage{multirow}
\usepackage{multicol}
\usepackage{float}
\usepackage{color}
\usepackage{amsopn}
\usepackage{mathrsfs}
\usepackage{mathtools}
\usepackage{amsmath}
\usepackage{arydshln}
\usepackage{blkarray}
\usepackage{courier}
\usepackage{rotating}
\usepackage{bm}
\usepackage{ragged2e}
\usepackage{setspace}
\usepackage{subfigure}
\usepackage{caption}

\usepackage{comment}
\usepackage{enumitem}
\usepackage{tabularx}
\usepackage{adjustbox}

\usepackage[linesnumbered,ruled,lined]{algorithm2e}
\usepackage{algorithmicx}
\usepackage{algpseudocode}
\usepackage{amsmath}

\definecolor{myblue}{rgb}{0, 0, 1}

\title{S$^3$: Social-network Simulation System with\\ Large Language Model-Empowered Agents}

\author{Chen Gao$^*$, Xiaochong Lan$^*$, Zhihong Lu, Jinzhu Mao, \\ \textbf{Jinghua Piao, Huandong Wang, Depeng Jin, Yong Li}\\
Department of Electronic Engineering,\\ Tsinghua University\\
\texttt{liyong07@tsinghua.edu.cn}\\
$^*$ These authors contributed equally to this work.
}

\begin{document}
\maketitle
\begin{abstract}
Simulation plays a crucial role in addressing various challenges within social science. It offers extensive applications such as state prediction, phenomena explanation, and policy-making support, among others. 
In this work, we harness the human-like capabilities of large language models (LLMs) in sensing, reasoning, and behaving, and utilize these qualities to construct the S$^3$ system (short for \textbf{S}ocial network \textbf{S}imulation \textbf{S}ystem).
Adhering to the widely employed agent-based simulation paradigm, we employ fine-tuning and prompt engineering techniques to ensure that the agent's behavior closely emulates that of a genuine human within the social network.
Specifically, we simulate three pivotal aspects: emotion, attitude, and interaction behaviors. By endowing the agent in the system with the ability to perceive the informational environment and emulate human actions, we observe the emergence of population-level phenomena, including the propagation of information, attitudes, and emotions.
We conduct an evaluation encompassing two levels of simulation, employing real-world social network data. Encouragingly, the results demonstrate promising accuracy.
This work represents an initial step in the realm of social network simulation empowered by LLM-based agents. We anticipate that our endeavors will serve as a source of inspiration for the development of simulation systems within, but not limited to, social science.
\end{abstract}

\section{Introduction}\label{sec::intro}

The social network, comprising interconnected individuals in society, constitutes a cornerstone of the contemporary world. 
Diverging from mathematical analysis, computer simulation offers a fresh avenue to comprehend the formation and evolution of social networks. This serves as a fundamental tool for social scientists.
Notably, in 1996, there was already a book titled \textit{Social Science Microsimulation}~\cite{troitzsch1996social} providing valuable insights about simulation from the perspective of social science. Social simulation encompasses a wide range of domains, encompassing both individual and population social activities.
At the heart of social simulation lie two perspectives~\cite{gilbert2005simulation}: 1) the dynamic feedback or interaction among individuals, and 2) the states of the population, either as a collective whole or as distinct groups. By simulating social activities, researchers and practitioners can predict the future evolution of individual and population states. In addition, they facilitate experimental environments through interventions.
Social simulation can be implemented in two forms: microlevel simulation~\cite{chopard1998cellular,park2023generative} and macrolevel simulation~\cite{kolesar1975simulation,meadows1974dynamics, forrester1993system,marsh1978using}. In macrolevel simulation, also known as system-based simulation, researchers model the dynamics of the system using equations that elucidate the changing status of the population. Conversely, microlevel simulation, or agent-based simulation, involves researchers employing either human-crafted rules or parameterized models to depict the behavior of individuals (referred to as agents) who interact with others.
Recently, with the exponential growth of the Internet, online social networks have emerged as the principal platform for societal activities. Users engage in various interactive behaviors such as chatting, posting, and sharing content. Consequently, the study of social networks has become a central research focus within the realm of social science, thereby emphasizing the criticality of simulation in this domain.

Large language models (LLMs)~\cite{brown2020language,openai2023gpt4,chowdhery2022palm, anil2023palm,touvron2023llama,zeng2022glm} are the recent advancement in the field of deep learning, characterized by the utilization of an extensive array of neural layers. These models undergo training on vast textual corpora, acquiring a remarkable fundamental capacity to comprehend, generate, and manipulate human language. 
Given their impressive prowess in text comprehension, which closely approximates human-level performance, LLMs have emerged as a particularly auspicious avenue of research for approaching general artificial intelligence. Consequently, researchers~\cite{aher2023using,horton2023large,hamalainen2023evaluating,park2023generative} leverage LLMs as agent-like entities for simulating human-like behavior, capitalizing on three fundamental capabilities.
First and foremost, LLMs possess the ability to perceive and apprehend the world, albeit restricted to environments that can be adequately described in textual form. 
Secondly, LLMs are capable of devising and organizing task schedules by leveraging reasoning techniques that incorporate both task requirements and the attendant rewards. 
Throughout this process, LLMs effectively maintain and update a memory inventory, employing appropriately guided prompts rooted in human-like reasoning patterns.
Lastly, LLMs exhibit the capacity to generate texts that bear a striking resemblance to human-produced language. These textual outputs can influence the environment and interact with other agents. Consequently, it holds significant promise to adopt an agent-based simulation paradigm that harnesses LLMs to simulate each user within a social network, thereby capturing their respective behaviors and the intricate interplay among users.

In this study, we present the Social-network Simulation System (S$^3$), which employs LLM-empowered agents to simulate users within a social network effectively. Initially, we establish an environment using real-world social network data. To ensure the authenticity of this environment, we propose a user-demographic inference module that combines prompt engineering with prompt tuning, to infer user demographics such as age, gender, and occupation. 
Within the constructed environment, users have the ability to observe content from individuals they follow, thereby influencing their own attitudes, emotions, and subsequent behaviors. Users can forward content, create new content, or remain inactive. Hence, at the individual level, we employ prompt engineering and prompt tuning methodologies to simulate attitudes, emotions, and behaviors. Notably, this simulation considers both demographics and memory of historically-posted content.
At the population level, the accumulation of individual behaviors, including content generation and forwarding, alongside the evolving internal states of attitudes and emotions, leads to the emergence of collective behavior. This behavior encompasses the propagation of information, attitudes, and emotions.

To assess the efficacy of the proposed S$^3$ system, we have chosen two exemplary scenarios, namely, \textbf{gender discrimination} and \textbf{nuclear energy}. With respect to gender discrimination, our objective is to simulate user responses to online content associated with this issue, while closely observing the dissemination patterns of related information and evolving public sentiment. Regarding nuclear energy, our aim is to simulate user reactions to online content pertaining to power policies. In addition, we aim to simulate the contentious and conflicting interactions between two opposing population groups. To evaluate the precision of our simulations, we employ metrics that measure accuracy at both the individual and population levels.
This work's main contributions can be summarized as follows.
\begin{itemize}[leftmargin=*]
    \item We take the pioneering step of simulating social networks with large language models (LLMs), which follows the agent-based simulation paradigm, and empowers the agents with the latest advances.
    \item We develop a simulation system that supports both individual-level and population-level simulations, which can learn from the collected real social network data, and simulate future states.
    \item We systematically conduct the evaluation, and the results show that the simulation system with LLM-empowered agents can achieve considerable accuracy in multiple metrics. Consequently, our system introduces a novel simulation paradigm in social science research, offering extensive support for scientific investigations and real-world applications.
\end{itemize}

To provide a comprehensive understanding of the current research landscape, we begin by reviewing relevant works in Section~\ref{sec::related}. Subsequently, we proceed to introduce the simulation system in Section~\ref{sec::system}, followed by a detailed exposition of the methodology and implementation in Section~\ref{sec::method}. In Section~\ref{sec::discussion}, we engage in discussions and analyze open challenges associated with related research and applications. Finally, we conclude our work in Section~\ref{sec::conclusion}.

\section{Related Works}\label{sec::related}
In this section, we discuss two areas close to this work, social simulation and large language model-based simulation.

\subsection{Social Simulation}

According to~\cite{bratley1987guide}, "Simulation means driving a model of a system with suitable inputs and observing the corresponding outputs". Social simulation aims to simulate various social activities, which encompass a wide range of applications~\cite{gilbert2005simulation}.
One primary advantage of social simulation is its potential to aid social scientists in comprehending the characteristics of the social world~\cite{axelrod1997advancing}.
This is primarily attributed to the fact that the internal mechanisms driving social behaviors are not directly observable.
By employing a simulation model capable of reasonably replicating the dynamic nature of historical social behaviors, it becomes feasible to utilize the simulation tool for predicting the future of the social system.
Furthermore, social simulation can serve as a training ground, particularly for economists involved in social-economic simulations~\cite{spencer1984effect}. In this context, the economist can assume a digital persona, namely an artificial intelligence program tasked with formulating economic policies.
Moreover, social simulation can even serve as a substitute for human presence, exemplified by the emergence of digital avatars in the metaverse~\cite{lee2021all}.
From the perspective of social science research, social simulation plays a crucial role in facilitating the development of new social science theories. It achieves this by validating theoretical assumptions and enhancing theory through the application of more precise formalizations.

In spite of the promising applications, conducting social simulation is complex. The earliest works use discrete event-based simulation~\cite{kolesar1975simulation} or system dynamics~\cite{meadows1974dynamics, forrester1993system,marsh1978using} with a series of equations to approximate multiple variables over time that partly describe the system.
These early methods primarily focused on accurately predicting the variables rather than elucidating the underlying mechanisms or causal relationships. 
Subsequently, drawing inspiration from the rapid development and remarkable success of simulation in other scientific domains, the utilization of agent-based simulation emerged in the field of social simulation. A notable and representative technique among these simulation methods is the employment of \textit{Cellular Automata}~\cite{chopard1998cellular}. Initially, this approach establishes a social environment composed of numerous individuals and subsequently formulates a set of rules dictating how individuals interact with one another and update their states.
Agent-based simulation can be regarded as a micro-level simulation that approximates real-world systems by describing the behavior of explicitly defined micro-level individuals. Thus, it is also referred to as microsimulation.

In recent times, owing to significant advancements in machine learning and artificial intelligence, agent-based simulation has witnessed a notable transformation. This transformation is characterized by the utilization of increasingly intricate and robust agents propelled by machine learning algorithms. These agents possess the ability to dynamically perceive their surroundings and exhibit actions that closely resemble human behavior.
The rapid progress in simulating individual agents has not only preserved the effectiveness of conventional simulation paradigms but has also resulted in significant improvements. This is particularly important for large language models, which are on the path towards achieving partial general artificial intelligence. Consequently, in this study, we embrace the microsimulation paradigm and employ meticulously guided and finely tuned large language models to govern the behavior of individuals within social networks.

\subsection{Large Language Model-based Simulation}

Recently, relying on the strong power in understanding and generating human language, large language models such as GPT series~\cite{brown2020language,openai2023gpt4}, PaLM series~\cite{chowdhery2022palm, anil2023palm},  LLaMA~\cite{touvron2023llama}, GLM~\cite{zeng2022glm}, etc. are attracting widespread attention.
LLMs have exhibited exceptional capabilities in zero-shot scenarios, enabling rapid adaptation to diverse tasks across academic and industrial domains. The expansive language model aligns well with the agent-based simulation paradigm mentioned earlier, wherein the primary objective involves constructing an agent represented by a rule or program endowed with sufficient capacity to simulate real-world individuals.

Aher~\textit{et al.}~\cite{aher2023using} conducted a preliminary test to find that LLMs possess the capability to reproduce some classic economic, psycholinguistic, and social psychology experiments.
Horton~\textit{et al.}~\cite{horton2023large} substitute human participants with LLM agents, which are given endowments, information, preferences, etc., with prompts and then simulate the economic behaviors.
The results with LLM-empowered agents show qualitatively similar results to the original papers (with human experiments)~\cite{samuelson1988status,charness2002understanding}.
Another study ~\cite{hamalainen2023evaluating} adopts an LLM-based crowdsourcing approach by gathering feedback from LLM avatars representing actual humans, to support the research of computational social science.

Recently, Part~\textit{et al.}~\cite{park2023generative} construct a virtual town with 25 LLM-empowered agents based on a video game environment, in which the agent can plan and schedule what to do in daily life. Although the simulation is purely based on a generative paradigm without any real-data evaluation, it provides insights that LLM can serve as a powerful tool in agent-based simulation. Each agent was assigned its own identity and distinct characteristics through prompts, facilitating communication among them. It is noteworthy that this simulation was conducted exclusively within a generative paradigm, without incorporating any real-world data for evaluation. Nevertheless, the findings offer valuable insights into LLM's potential as a potent tool in agent-based simulations.

\section{S$^3$: Social Network Simulation }\label{sec::system}

\subsection{System Overview}

Our system is constructed within a social network framework, wherein the agent's capabilities are augmented through the utilization of large language models. More specifically, our primary objective is to ensure that the simulation attains a significant degree of quantitative accuracy, catering to both individual-level and population-level simulations. Regarding individual-level simulation, our aim is to replicate behaviors, attitudes, and emotions by leveraging user characteristics, the informational context within social networks, and the intricate mechanisms governing user cognitive perception and decision-making. Through the utilization of agent-based simulation, we further assess the population-level dynamics by scrutinizing the performance of simulating three pivotal social phenomena: the propagation process of information, attitude, and emotion.

\begin{figure}[t!]
	\centering
	\includegraphics[width=0.85\textwidth]{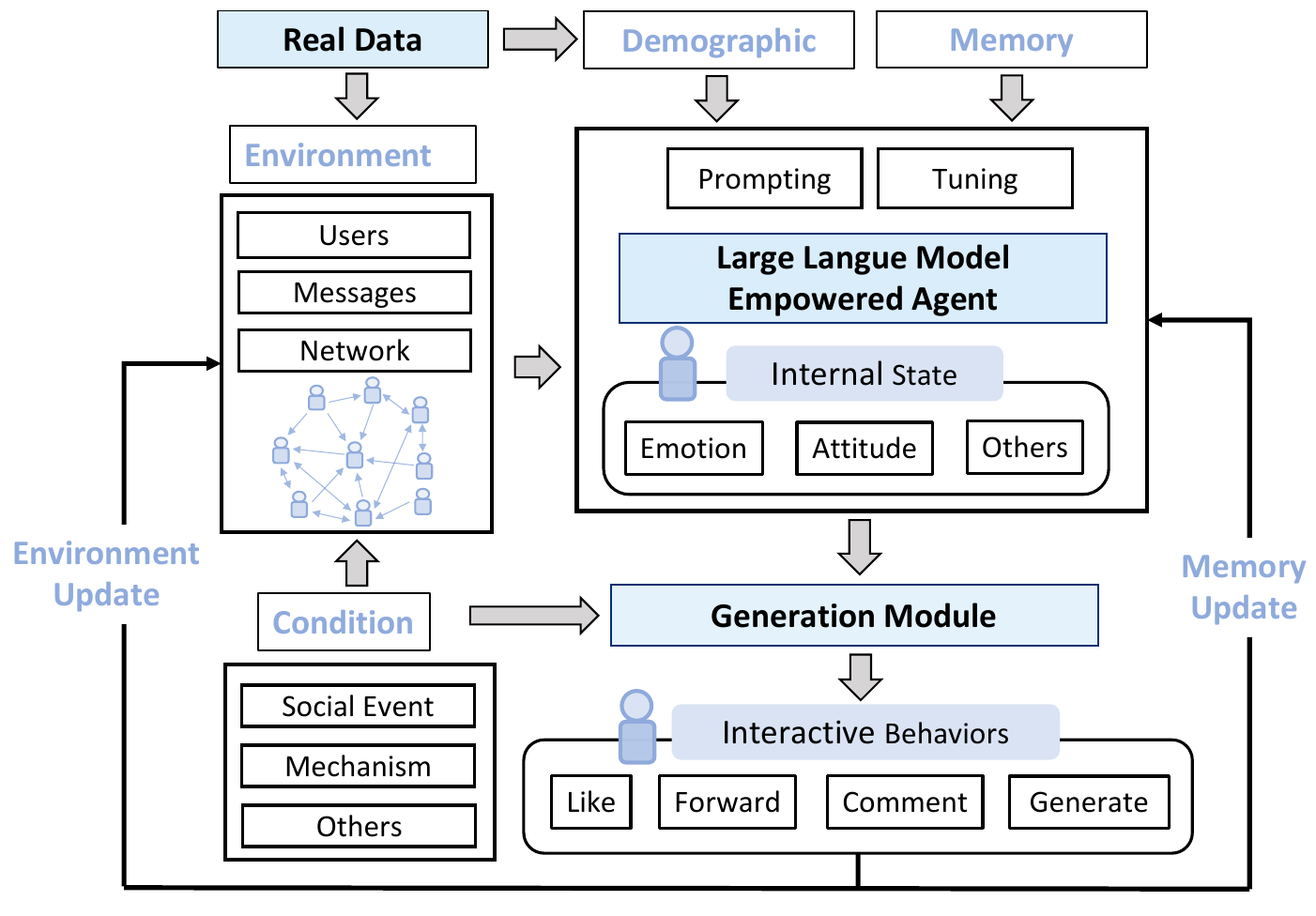}
	\caption{The overview of the social network simulation system.}
	\label{fig::framework}
\end{figure}

\subsection{Social Network Environment}
\begin{table}[t!]
	\caption{The utilized datasets for social network simulation.}
    \label{tab::overall}
    \vspace{0.2cm}
	\centering
	\begin{tabular}{lccc}
		\toprule
		\textbf{Scenario} & \textbf{\#Users} & \textbf{\#Relations} & \textbf{\#Posts} \\ 
		\midrule
		Gender Discrimination & 8,563 & 25,656 & 103,905 \\ 
		Nuclear Energy & 17,945 & 77,435 & 229,450 \\ 
		\bottomrule
	\end{tabular}
\end{table}

In this study, our focus is directed toward two specific focal points, namely gender discrimination and nuclear energy. These particular subjects are chosen owing to their highly controversial nature, which yielded an extensive corpus of data. More specifically, our investigation regarding nuclear energy centers on examining the prevailing attitudes of the general public toward the choice between supporting nuclear energy sources or relying on fossil fuels. As for gender discrimination, our objective is to delve into the emotional experiences of individuals and populations, particularly those elicited by incidents of gender-based discrimination, such as feelings of anger. The availability of such copious amounts of data facilitates the extraction of a substantial portion of the authentic network, thereby enabling us to gain a macroscopic perspective that closely approximates reality. To conduct this analysis, we collect the real data with users, social connections, and textual posts in social media, as detailed in Table \ref{tab::overall}. This dataset provides us with the necessary resources to delve deep into the dynamics of these contentious subjects and gain valuable insights into their impact on social networks. 

User demographics play a pivotal role in shaping user behavior, necessitating the development of a more extensive user persona to enable the realistic and plausible simulation of their actions. However, due to the limited availability of user information obtained directly from social media, it becomes imperative to extract the missing user demographics from textual data, such as user posts and personal descriptions. 
Specifically, we capture user demographic features from textual information using LLM, with a particular emphasis on predicting Age, Gender, and Occupation. By integrating demographic attributes inferred from social network data, we are able to present an enhanced and more authentic representation of users' actions and interactions.

\begin{table}[t!]
\centering
\caption{Performance of our system on prediction tasks for individual simulation.}
\vspace{0.2cm}
\label{tab::ind1}
\small
\begin{tabular}{llccc}
\toprule
\textbf{Scenario} & \textbf{Task} & \textbf{Acc} & \textbf{AUC} & \textbf{F1} \\ 
\midrule
\multirow{2}{*}{Gender Discrimination} & Emotion Level & 71.8\% & --- & --- \\
                                 & Event Propogation & 66.2\% & 0.662 & 0.667 \\
\midrule
\multirow{3}{*}{Nuclear Energy}  & Initial Attitude & 74.3\% & 0.727 & 0.834 \\ 
                                 & Attitude Change & 83.9\% & 0.865 & 0.857 \\
                                 & Event Propogation & 69.5\% & 0.681 & 0.758 \\ 
\bottomrule
\end{tabular}
\end{table}

\subsection{Individual-level Simulation}

Utilizing the initialized social network environment, the system commences the simulation at an individual level. Precisely, the user acquires awareness of the information environment, thereby influencing their emotions and attitude. Subsequently, the user is granted the option to forward (repost) observed posts, generate new content, or keep inactive. In essence, we conduct individual simulations encompassing three facets: emotion, attitude, and interaction behavior.

\subsubsection{Emotion Simulation}
In the process of disseminating real-world events, when a user with their own cognition, attitudes, and personality encounters an event, they are often triggered emotionally and express their emotions on social platforms. Emulating user emotions is crucial for social network simulations, as it significantly influences how users convey their intended messages. However, simulating emotions is challenging due to the multitude of factors and complex relationships involved in human emotions. Leveraging the rich knowledge of human behavior inherent in LLMs, we employ LLM to simulate individual emotions.

Specifically, we model the potential emotions of users towards a particular event as three levels: calm, moderate, and intense. Initially, when users are unaware of the event, their default emotion level is set to calm. However, as they become aware of the event, their emotional state begins to evolve. In order to capture this dynamic nature of emotions, we employ a Markov process. This process considers several factors, including the user's current emotion level, user profiles, user history, and the messages received at the present time step. By integrating these variables, we can predict the user's emotion level in the subsequent time step.

Our emotion simulation approach has yielded promising results at the individual level. As shown in Table~\ref{tab::ind1}, using real-world data for evaluation, our method demonstrates good performance in predicting the emotions of the next time step. We achieve an accuracy of 71.8\% in this three-classification task, thanks to the excellent modeling and understanding of human emotional expression by large language models.

\subsubsection{Attitude Simulation}
Just as emulating user emotions proves pivotal for social network simulations, simulating user attitudes carries equal weight. The reproduction of attitudes in a virtual social environment is complex yet indispensable. It is the combination of these attitudes that guide users' actions, opinions, and decisions about different topics. The challenge in this simulation lies in the multifaceted and subjective nature of attitudes, which are influenced by a wide range of internal and external factors, from individual experiences and beliefs to societal influences and perceived norms.

For our simulation, we assume that users have initial attitudes towards specific issues, which change based on unfolding events. This dynamic adaptation of attitudes is reflective of real-world social interactions, where people modify their views in response to changing circumstances, influential figures, or compelling arguments. 

In our model, much akin to the emotional state, we track the users' attitudes on a binary spectrum, which consists only of negative and positive stances towards an event. Our first step is to establish an initial state for the user's attitude. This is derived from the user profiles and user history, reflecting their predispositions based on past interactions and behaviors. Once the initial state is established, the dynamics of attitude changes are modeled as a Markov process. The subsequent evolution of these attitudes incorporates not only the user's current attitude but also their profile, history, and the messages received at the current time step. These factors are collectively employed to predict the user's attitude in the ensuing time step. Both the initial attitude and the assessment of attitude change are determined based on the LLM.

As depicted in Table~\ref{tab::ind1}, our methods have demonstrated excellent performance. In the task of predicting initial attitudes, our approach yields an accuracy of 74.3\%, an AUC score of 0.727, and an F1-Score of 0.834. In the subsequent task of attitude change prediction, our method performs even better, achieving an impressive accuracy of 83.9\%, an AUC score of 0.865, and an F1-Score of 0.857. These results can be largely attributed to the ability of LLMs to profoundly comprehend human behavior and cognition. Such understanding enables a sophisticated interpretation of user-generated content, resulting in a more accurate prediction of users' attitudes and their evolution over time.

\begin{table}[t!]
\centering
\caption{Performance of our system on conditional text generation tasks.}
\vspace{0.3cm}
\label{tab:ind2}
\begin{tabular}{ccc}
\hline
\textbf{Scenario}              & \textbf{Perplexity} & \textbf{Cosine Similarity} \\ \hline
\textbf{Gender Discrimination} & 19.289              & 0.723                              \\ \hline
\textbf{Nuclear Energy}        & 16.145              & 0.741                              \\ \hline
\end{tabular}
\end{table}

\subsubsection{Content-generation Behavior Simulation}
Within the realm of real-world social networks, users shape their content based on their prevailing attitudes and emotions towards distinct events. Emulating this content creation process is an essential, yet complex, aspect of social network simulations. Each piece of generated content acts as a mirror to the user's internal state and external influences, manifesting their individual perspective on the event at hand. The crux of the challenge is to encapsulate the wide array of expressions and styles that users employ to convey their sentiments, opinions, and reactions.

Leveraging the strengths of LLMs can significantly alleviate this challenge. These models, with their ability to generate text that closely resembles human-like language patterns, facilitate the simulation of user-generated content with high accuracy. By inputting the user's profile, along with their current attitude or emotional state, these models are capable of generating content that faithfully reproduces what a user might post in response to a particular event.

This approach, informed by the capabilities of large language models, enables us to craft a sophisticated simulation that mirrors the content generation process in real-world social networks. It thereby provides a nuanced understanding of how users' attitudes and emotions are reflected in their content, offering invaluable insights for the study of social dynamics.

As can be seen in Table~\ref{tab:ind2}, our methods yield impressive results. In the Gender Discrimination scenario, we achieved a Perplexity score of 19.289 and an average cosine similarity of 0.723 when compared with the actual user-generated text. In the case of the Nuclear Energy scenario, these figures were even more impressive, with a Perplexity score of 16.145 and an average cosine similarity of 0.741.

These results validate the effectiveness of our approach, where the LLM's profound comprehension of human cognition and behavior significantly contributes to accurately simulating user-generated content in social network simulations. Thus, our model serves as a powerful tool in understanding and predicting social dynamics in various contexts.

\subsubsection{Interactive Behavior Simulation}

During the simulation, upon receiving a message from one of their followees, the user is faced with a consequential decision: whether to engage in forwarding, posting new content or do nothing. 
Effectively modeling the decision-making process is important in simulating information propagation.

Through our data-driven approach, we utilize Large Language Models (LLMs) to simulate users' interaction behavior by capturing the intricate relationship between users and contexts. The input is the information environment that the user senses, and the LLM-empowered agent make the decision by learning from the observed real data. 

Our model has demonstrated commendable efficacy in this regard. In the scenario of Gender Discrimination, our model achieved an Accuracy of 66.2\%, AUC of 0.662, and F1-Score of 0.667. Progressing to the Nuclear Energy context, the model's performance remained robust, with an Accuracy of 69.5\%, AUC of 0.681, and F1-Score of 0.758.

These promising results not only attest to the LLM's capability in accurately simulating individual user behavior but also pave the way for exploring its potential at a larger scale. This accomplishment forms the basis for the population-level simulation, which we will delve into in the subsequent sections.

\subsection{Population-level Simulation}

In S$^3$, we capture three forms of propagation, including the propagation of information, emotion, and attitude. Here information propagation focuses on the transmission of news that describes events in social environments. Emotion propagation emphasizes the social contagion of people's feelings toward specific events or topics. Attitude propagation describes that people exchange their attitudes or viewpoints in the social network. Subsequently, we shall expound upon our comprehensive capacity to simulate these three aforementioned forms of propagation.

\begin{figure*}[]
\centering
\subfigure[True spread]{               
\includegraphics[width=3.3cm]{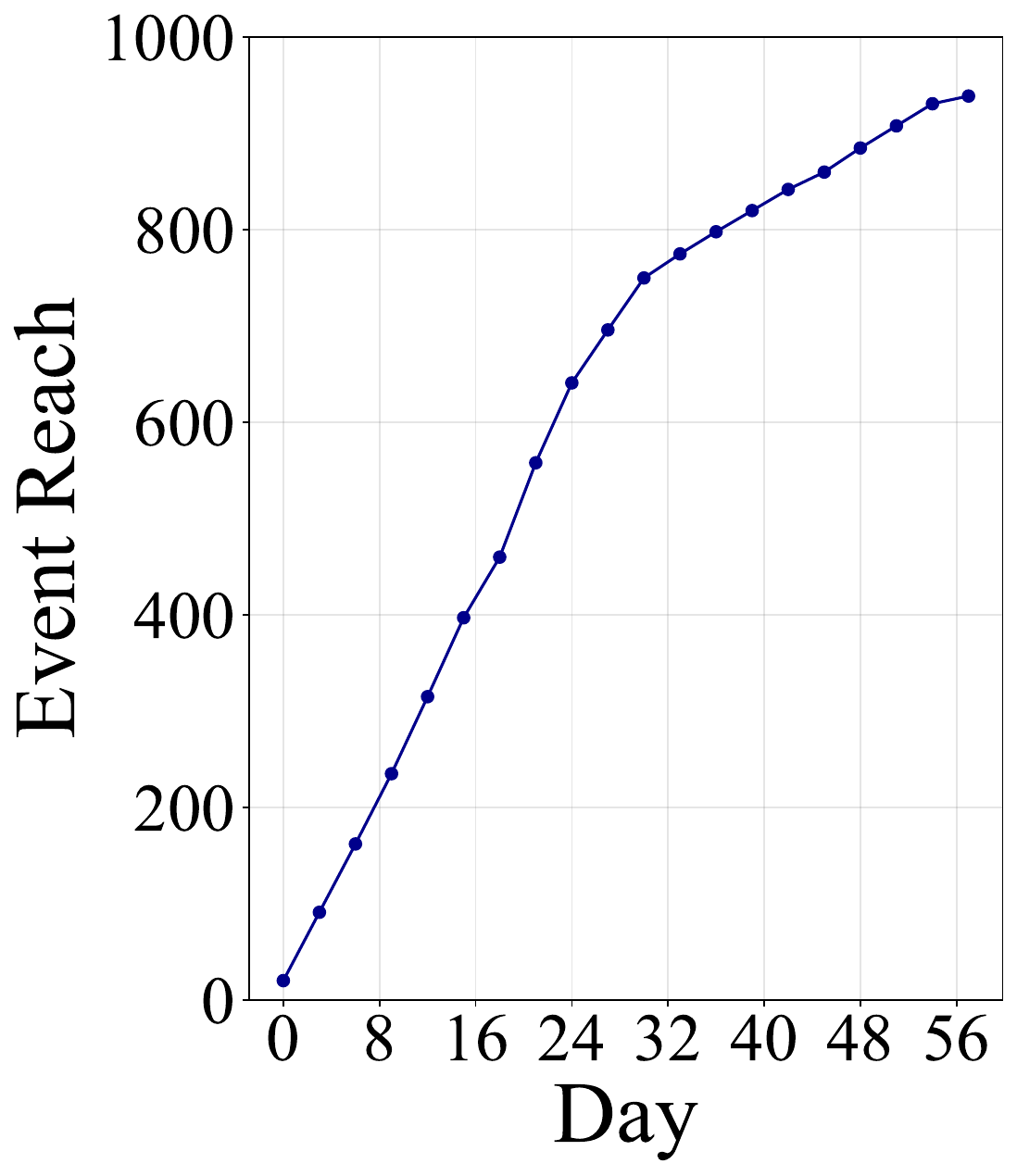}\label{fig::tsb}}
\subfigure[Simulated spread]{
\includegraphics[width=3.3cm]{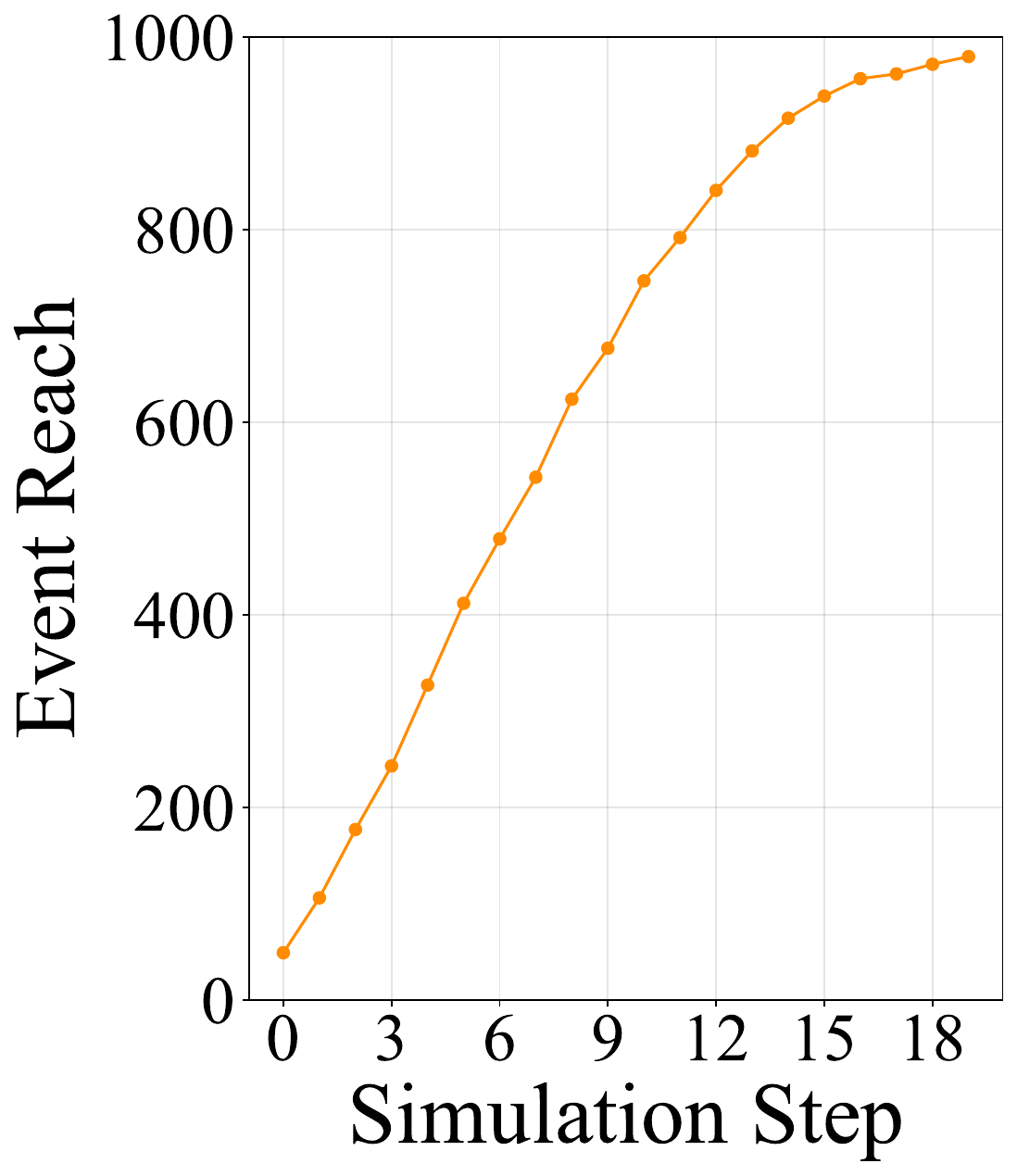}\label{fig::ssb}}
\subfigure[True emotion trend]{
\includegraphics[width=3.3cm]{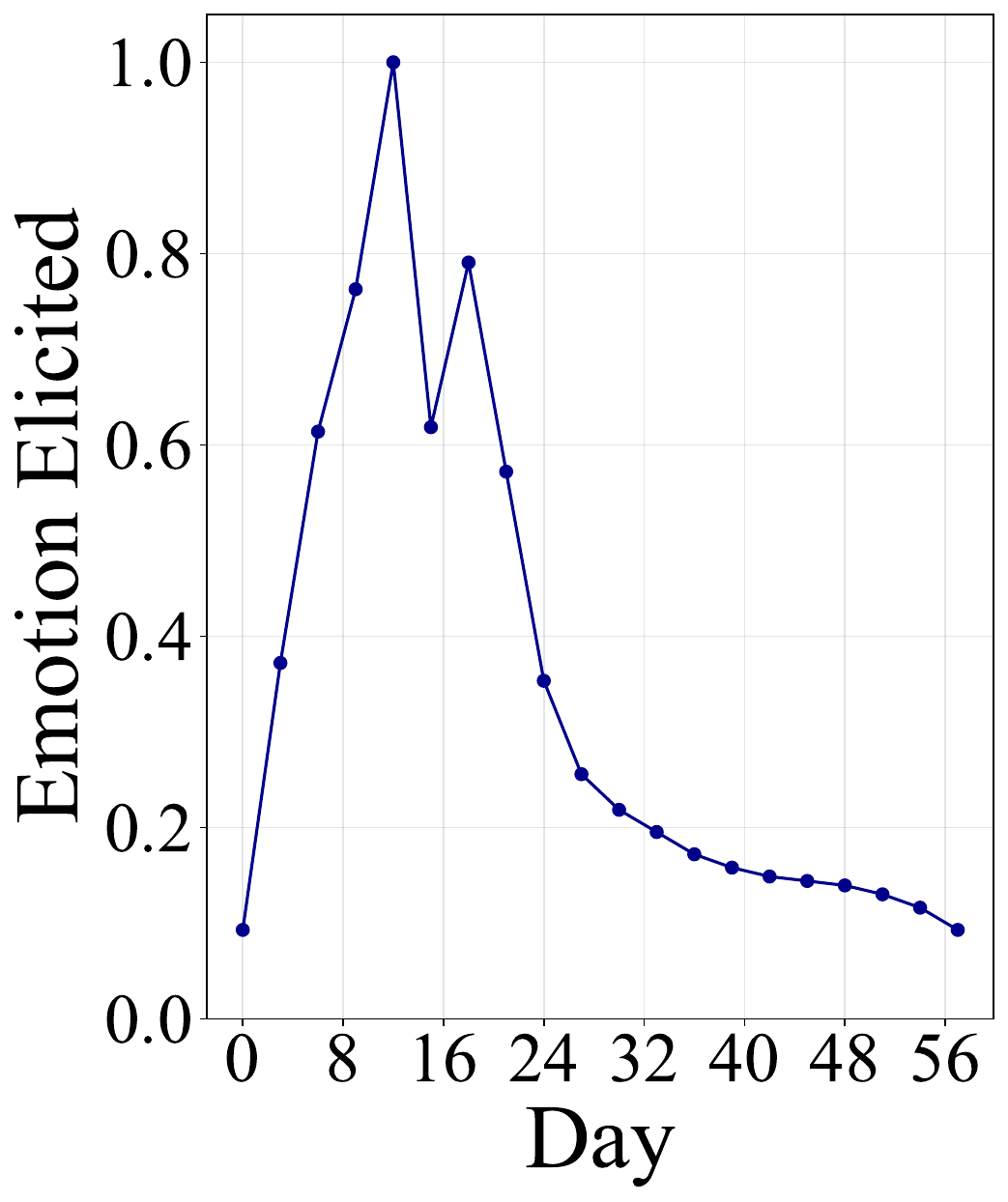}\label{fig::teb}}
\subfigure[Simulated emotion trend]{
\includegraphics[width=3.3cm]{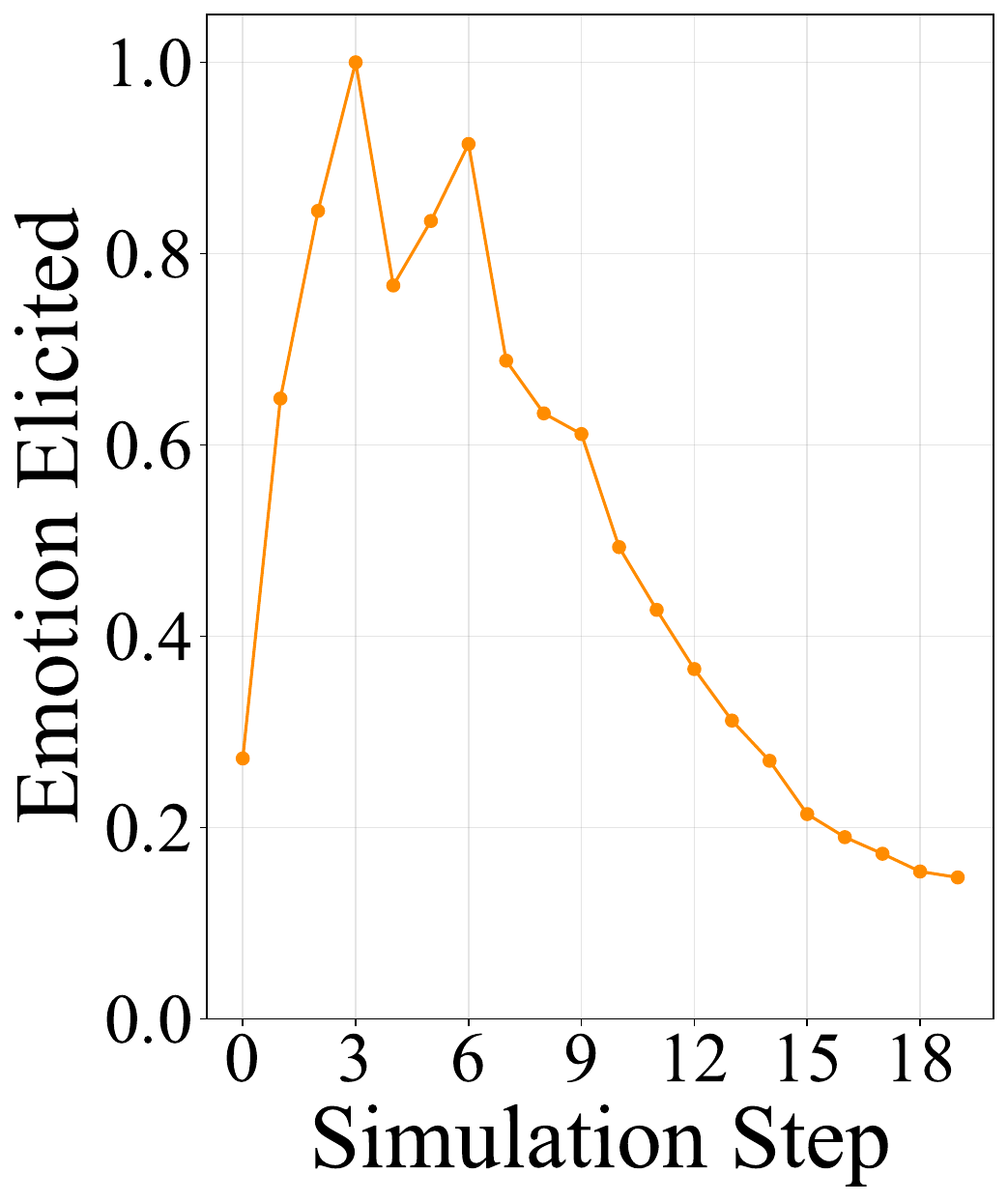}\label{fig::seb}}

\caption{True spread, simulated spread, true emotion trend and simulated emotion trend of Eight-child Mother Event.}
\label{fig::sb}
\end{figure*}

\begin{figure*}[]
\centering
\subfigure[True spread]{               
\includegraphics[width=3.3cm]{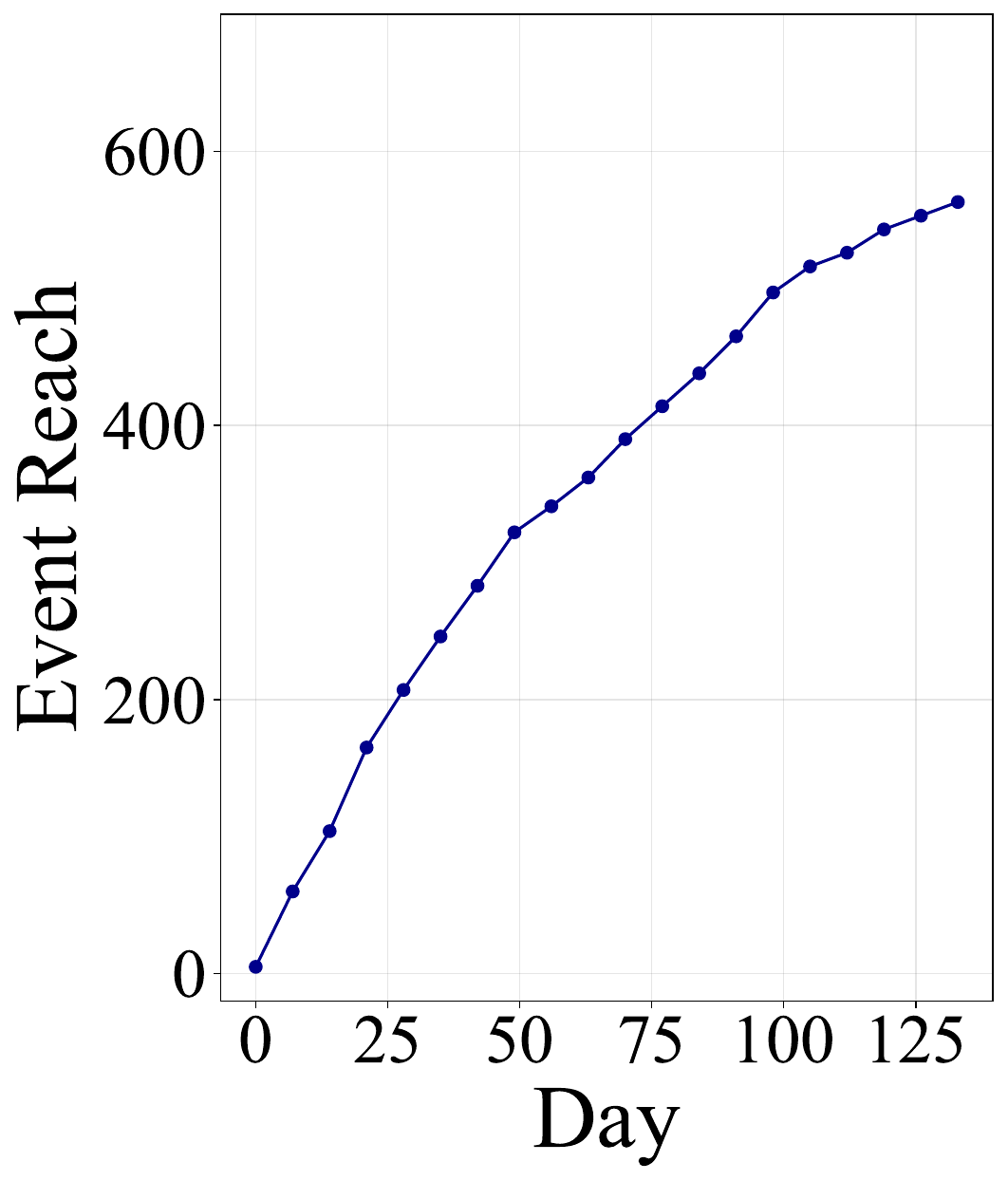}\label{fig::tsn}}
\subfigure[Simulated spread]{
\includegraphics[width=3.3cm]{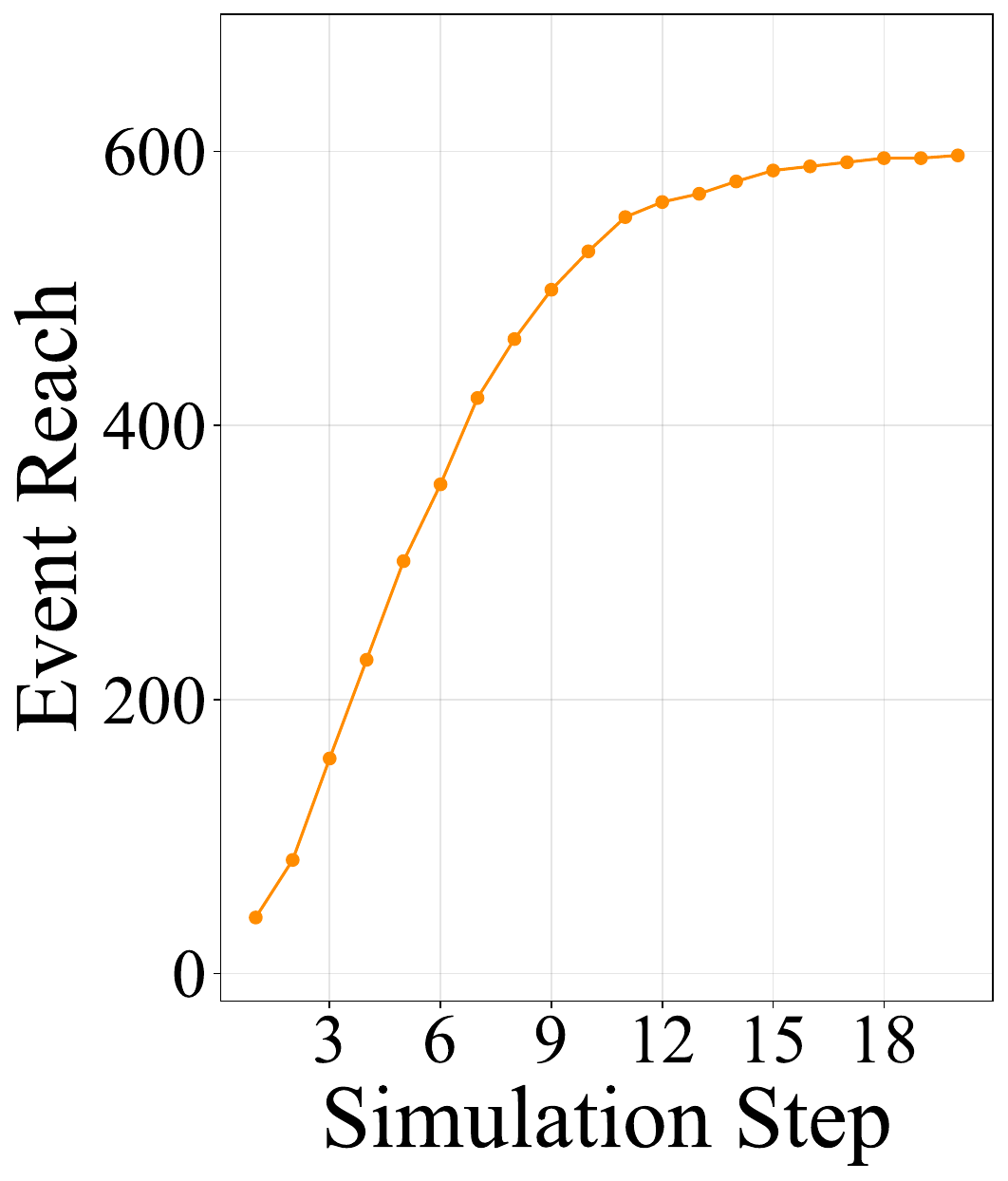}\label{fig::ssn}}
\subfigure[True change of attitudes]{               
\includegraphics[width=3.3cm]{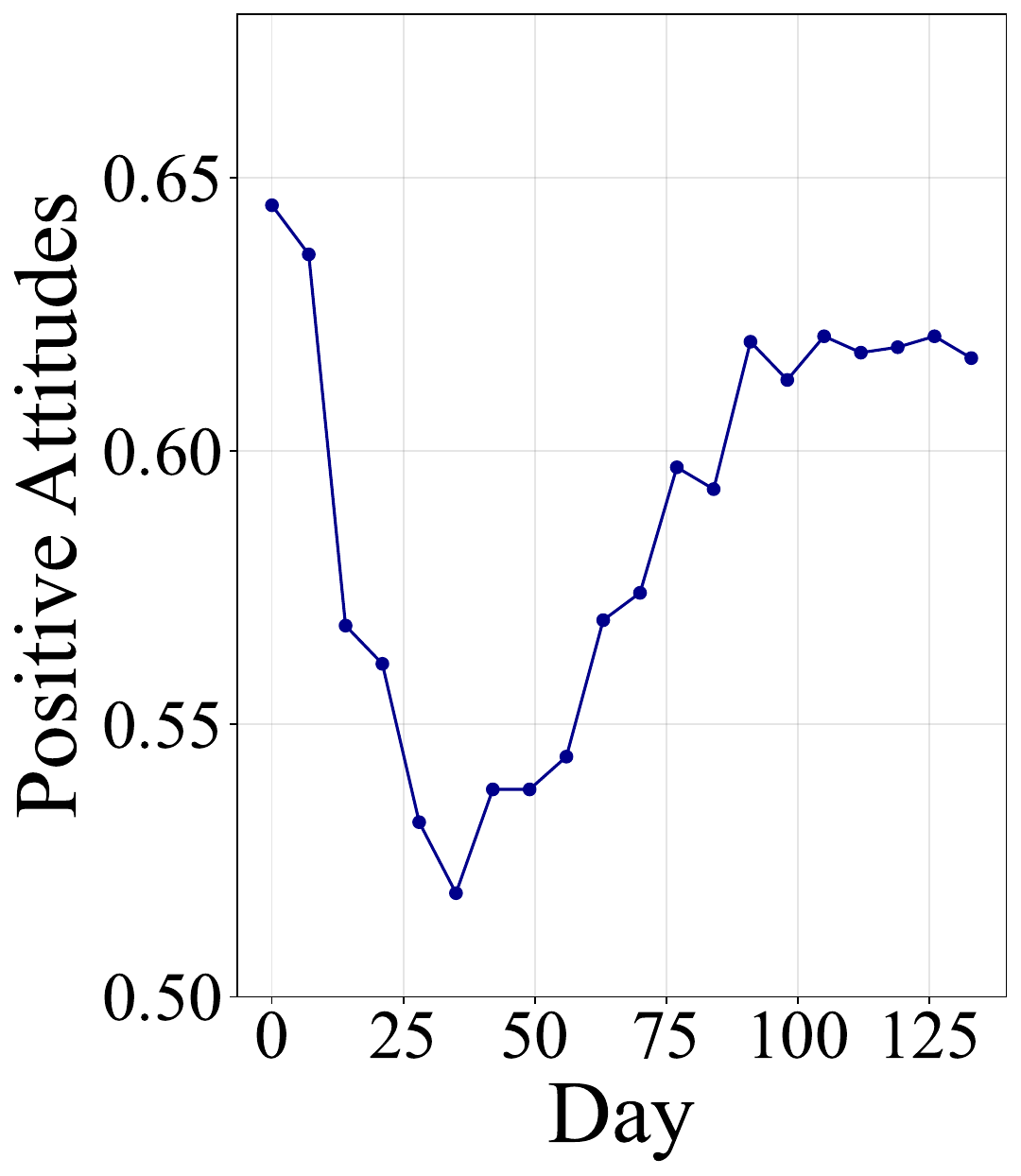}\label{fig::tpn}}
\subfigure[Simulated change of attitudes]{
\includegraphics[width=3.3cm]{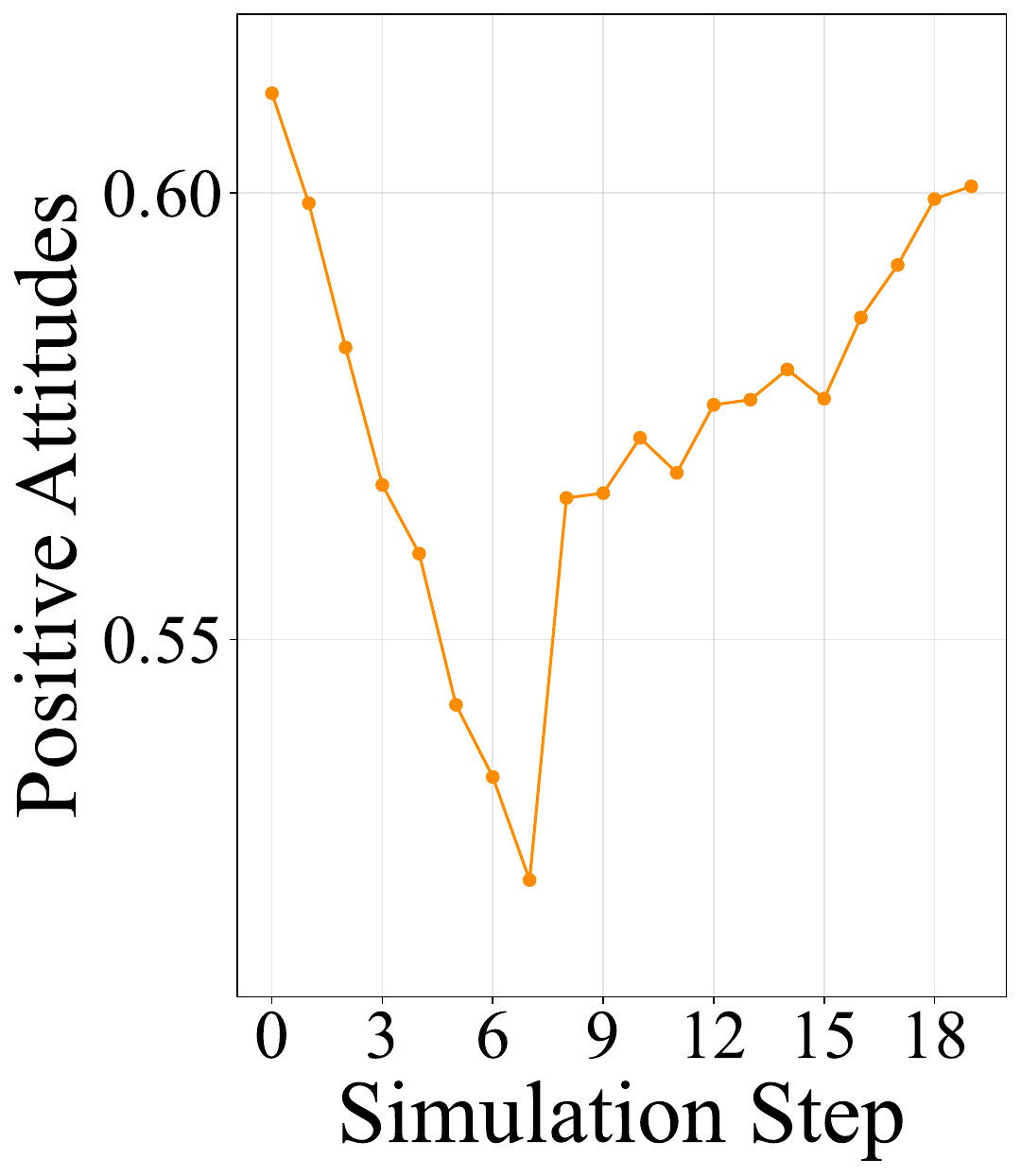}\label{fig::spn}}
\caption{True spread, simulated spread, true and simulated changes in proportion of positive attitudes towards nuclear energy during the Japan Nuclear Waste Water Release Event.}
\label{fig::pn}
\end{figure*}

\subsubsection{Information Propagation}

With the widespread adoption of digital media, the propagation of information experiences a significant acceleration~\cite{lorenz2023systematic,luding2005information}. In the context of a simulation system designed to mimic social networks, one of its paramount functionalities lies in accurately modeling the process of information propagation and delineating crucial phase transitions~\cite{xie2021detecting,notarmuzi2022universality}. For example, Notarmuzi et al.~\cite{notarmuzi2022universality} conducted extensive empirical studies on a large scale, successfully distilling the concepts of universality, criticality, and complexity associated with information propagation in social media. Meanwhile,  Xie et al.~\cite{xie2021detecting} expanded upon the widely accepted percolation theory and skillfully captured the intricate phase transitions inherent in the spread of information on social media platforms.

Diverging from previous studies grounded in physical models, our approach adopts a LLM perspective to capture the dynamics of the information propagation process. In order to ascertain the efficacy of our proposed S$^3$ model, we have selected two typical events: (i) Eight-child Mother Event and (ii) Japan Nuclear Wastewater Release Event. The former event came to public attention in late January 2022, encompassing a range of contentious issues, such as sexual assault and feminism. The latter event entails Japan's government's decision to release nuclear wastewater into the ocean, eliciting significant global scrutiny and interest.

Utilizing our simulator as a foundation, we employ a quantitative approach to evaluate the temporal dissemination of the aforementioned occurrence. This is achieved by calculating the overall number of people who have known the events at each time step  (refer to Figure~\ref{fig::ssb} and Figure~\ref{fig::ssn}). Subsequently, through a comparative analysis with the empirical data (as illustrated in Figure~\ref{fig::tsb} and Figure~\ref{fig::tsn}), we discern that our simulator exhibits a commendable capacity for accurately forecasting the propagation patterns of both events. In particular, we notice that the rate of rise becomes gradually marginal over time, which can also be captured by our simulator.

\subsubsection{Emotion Propagation}

Another indispensable form of propagation is the transmission of emotion on social media~\cite{wang2022global,schafer2002spinning}. For example, Wang et al.~\cite{wang2022global} adopt the natural language processing techniques (BERT) and perform frequent global measurements of emotion states to gauge the impacts of pandemic and related policies. In S$^3$, we utilize the state-of-the-art LLM to extract emotions from real-world data and simulate the emotional propagation among LLM-based agents. 

To examine whether the S$^3$ simulator can also reproduce the emotion propagation process, we further simulate users' emotions expressed in the Eight-child Mother event. We extract the emotional density from the textual interactions among agents. Comparing our simulation results (Figure~\ref{fig::seb}) and the empirical observations (Figure~\ref{fig::teb}),  we find that our model can well capture the dynamic process of emotion propagation. Notably, we observe that there are two emotional peaks in the event. This suggests that if news of the event spreads more slowly across a larger community, a secondary peak in emotional intensity may occur. Based on the initialization obtained from real-world data, our model successfully reproduces these distinct peaks, thereby demonstrating  the effectiveness of our proposed S$^3$ system.

\subsubsection{Attitude Propagation}
One of today's most concerning issues is the polarization and confrontation between populations with diverging attitudes toward controversial topics or events. Great efforts have been made to quantify real-world polarization~\cite{lorenz2023systematic,flamino2023political,hohmann2023quantifying} and simulate the polarization process using co-evolution model~\cite{santos2021link,baumann2020modeling,baumann2021emergence,liu2023emergence}. In S$^3$, we use LLM to simulate propagation attitudes and predict polarization patterns in social networks.

Here we focus on the Japan Nuclear Wastewater Release Event, in which people's attitudes are polarized toward nuclear energy. As shown in Figure~\ref{fig::pn}, we can observe that with the propagation of related information, positive attitudes toward nuclear energy decline rapidly, exhibiting a salient trough. In our S$^3$ model, though modeling repeated interactions among agents, we reproduce the sudden decrease in positive attitudes and also capture their gradual increase. Overall, these observations suggest that our proposed model can not only simulate attitude propagation but also capture the critical dynamical patterns when situated in real-world scenarios.

\subsection{Comparative Evaluation}

To comprehensively evaluate the effectiveness of our S$^3$ system, we conduct comparative experiments against several baseline methods across three key propagation tasks: information propagation, opinion propagation, and emotion propagation.

\subsubsection{Information Propagation Comparison}

We compare our approach with two widely-used baseline models: the Linear Threshold (LT) model~\cite{chen2010scalable} and the Independent Cascade (IC) model~\cite{he2014minimum}. We employ two evaluation metrics: Mean Squared Error Difference (MSED) and Correlation Coefficient (Cor). MSED is defined as:

\begin{equation}
\text{MSED} = \frac{1}{T} \sum_{t=1}^{T} (S_t^{actual} - S_t^{predicted})^2
\end{equation}

where $S_t^{actual}$ and $S_t^{predicted}$ are the actual and predicted state values at time $t$, respectively.

\begin{table}[t!]
\centering
\caption{Performance comparison for information propagation simulation.}
\vspace{0.2cm}
\label{tab:info_prop}
\begin{tabular}{lccc}
\hline
\textbf{Method} & \textbf{Dataset} & \textbf{MSED} & \textbf{Cor} \\
\hline
LT & Nuclear Energy & 0.081 & 0.965 \\
IC & Nuclear Energy & 0.075 & 0.971 \\
S$^3$ & Nuclear Energy & 0.103 & 0.967 \\
\hline
LT & Gender Discrimination & 0.065 & 0.982 \\
IC & Gender Discrimination & 0.074 & 0.966 \\
S$^3$ & Gender Discrimination & 0.051 & 0.996 \\
\hline
\end{tabular}
\end{table}

As shown in Table~\ref{tab:info_prop}, our method achieves competitive performance compared to traditional models, with particularly strong results on the Gender Discrimination dataset where we achieve the lowest MSED and highest correlation.

\subsubsection{Opinion and Emotion Propagation Comparison}

For opinion and emotion propagation, we compare against five baseline methods: Voter~\cite{yildiz2010voting}, DeGroot~\cite{degroot1974reaching}, Feed-forward Neural Network (FNN), Sociologically-Informed Neural Network (SINN)~\cite{okawa2022predicting}, and Neural Dynamics on Complex Networks (NDCN)~\cite{zang2020neural}. These baseline methods require training data (50\% for training, 10\% for validation, 40\% for evaluation), while our approach operates in a zero-shot manner.

\begin{table}[t!]
\centering
\caption{Performance comparison for opinion and emotion propagation simulation.}
\vspace{0.2cm}
\label{tab:opinion_emotion}
\begin{tabular}{lcccc}
\hline
\multirow{2}{*}{\textbf{Method}} & \multicolumn{2}{c}{\textbf{Opinion Propagation}} & \multicolumn{2}{c}{\textbf{Emotion Propagation}} \\
\cline{2-3} \cline{4-5}
& \textbf{MSED} & \textbf{Cor} & \textbf{MSED} & \textbf{Cor} \\
\hline
Voter & 0.725 & -0.01 & 0.892 & -0.01 \\
DeGroot & 5.614 & 0.373 & 6.763 & 0.242 \\
FNN & 1.150 & 0.629 & 0.381 & 0.583 \\
SINN & 0.163 & 0.892 & 0.188 & 0.794 \\
NDCN & 0.060 & 0.882 & 0.060 & 0.828 \\
S$^3$ (Zero-shot) & 0.182 & 0.858 & 0.051 & 0.892 \\
\hline
\end{tabular}
\end{table}

Table~\ref{tab:opinion_emotion} demonstrates that our zero-shot approach achieves competitive performance with training-dependent baselines for opinion propagation and outperforms all baselines for emotion propagation, highlighting the effectiveness of our LLM-based approach.

\section{Architecture and Methodology}\label{sec::method}

\subsection{Architecture Design}

In order to simulate the process of information propagation on the online social network, we have designed a message propagation simulation framework illustrated in Figure \ref{fig::framework} and is explained in detail below.

\textbf{Environment Construction}: The construction of the environment involves the formation of a social network on a public platform, comprising users and connections among them. For instance, users have the ability to establish mutual following relationships with their friends, or one-way following relationships with users they find interesting. Hence, the social network can be characterized as a directed graph, where the outdegree and indegree of nodes in the network represent the number of people they follow and the number of followers they possess, respectively.
The users within this network can be broadly categorized into three groups: influential users, regular users, and low-impact users. Influential users typically exhibit a significantly larger number of followers compared to the number of people they follow. Moreover, they demonstrate a tendency to share high-quality original information. Regular users, on the other hand, typically maintain a balanced proportion of followers and followings. Additionally, a considerable portion of regular users engage in mutual following relationships, which often reflect their real-life friendships. Conversely, low-impact users exhibit limited followers, infrequent message posting, and typically represent the terminal points of message propagation chains. It is important to note that within this framework, we have excluded the consideration of social bots and zombie users, despite their prevalence on social platforms.

\textbf{User Characterization}
In addition to the social relationships present within the network, each user possesses their own attribute descriptions. Certain attributes are objective and specific, encompassing factors such as gender, occupation, and age. On the other hand, other attributes are more abstract, including their attitudes towards specific events and their prevailing emotional states. The former attributes tend to exhibit minimal fluctuations over short durations, whereas the latter attributes are more dynamic, particularly when users engage in information browsing on social platforms. In such cases, their fundamental attributes, message content, and message sources consistently shape their attitudes, emotions, and other abstract attributes. 
In light of the aforementioned descriptions, we also introduce a memory pool for each user. Given the abundance of messages from diverse users on online public platforms, a multitude of messages emerge daily. It is important to acknowledge that different messages exert varying influences on distinct users. To address this, we draw inspiration from~\cite{park2023generative} and propose the concept of influence factors. These factors calculate weighted scores based on parameters such as posting time, content relevance, and message importance. By doing so, we ensure that the user's memory pool retains the most impactful messages, making them highly memorable.

\begin{itemize}[leftmargin=*]
    \item \textbf{Temporal Influence}: The recency of messages plays a significant role in human memory, with previous messages gradually fading over time. A time score is ascribed to messages using a prescribed forgetting function.
    
    \item \textbf{Content Relevance}: The relevance of message content is assessed with regard to the user's individual characteristics. Notably, younger individuals tend to exhibit a greater inclination towards entertainment-related events, whereas middle-aged individuals demonstrate heightened interest in political affairs. To quantify the degree of relevance, a relevance score is obtained by measuring the cosine similarity between a user's fundamental attributes and the content of the message.
    
    \item \textbf{Message Authenticity}: The authenticity of messages is closely related to their sources. Messages are categorized based on their origins, encompassing messages disseminated by unidirectional followers, messages shared by mutual followers, messages recommended by the platform, and messages previously posted by the user themselves. Distinct scores are assigned to messages based on their respective sources.
\end{itemize}

\textbf{Update and Evolution Mechanism}: During a social gathering, various official accounts and individual users contribute posts concerning the event, encompassing news reports and personal viewpoints. Upon encountering these messages, the users who follow them manifest diverse emotional responses. Some users may even formulate their own stances on contentious matters, either in support or opposition, subsequently engaging in online activities such as endorsing, disseminating, and creating original message. In this simulation, we employ large language models to replicate individual users, leveraging their profiles and memory pools as prompts to generate cognitive reactions and behavioral responses. Subsequently, their abstract attributes and memory pools undergo updates. Following the modification of a user's memory pool, these messages disseminate and exert influence on their followers while they peruse the content. This iterative process persists, emulating the propagation of messages and the evolution of individuals' cognitive states.

\subsection{Initialization}

\subsubsection{Social Network Construction}

Within the scope of this study, we propose an initialization approach to construct a network utilizing data acquired from real-world social media sources (refer to Table \ref{tab::overall}). Strict adherence to privacy regulations and policies is maintained throughout the collection of social media data. Our approach leverages keyword-matching techniques to effectively extract posts relevant to the simulated scenarios. Subsequently, we delve into the identification of the authors and extract them as the foundational nodes of our network. Expanding beyond the individual level, we meticulously gather socially connected users. To establish connections between users, directed edges are established if the corresponding followee exists within the extracted user set. To optimize simulation efficiency, in this work, we focus solely on this sub-graph rather than the entire graph which is too large. During the simulation, the dissemination of messages occurs exclusively between source nodes and their corresponding target nodes.

\subsubsection{User Demographics Prediction}

Expanding upon the properties of the node, specifically focusing on user demographic attributes, emerges as a pivotal stride in our endeavor towards a more exhaustive simulation. Through the incorporation of additional information regarding the users into the system, we can delve into and scrutinize their behaviors, interactions, and influence within the network, more effectively. User demographic attributes allow us to capture heterogeneity and diversity in real-world social networks. That is, demographic attributes play a significant role in shaping individual behaviors and preferences, which, in turn, influence the network's overall attitude dynamics. In our study, we chose gender, age, and occupation as the major demographic attributes. As social media data does not directly offer attributes such as gender, age, and occupation, we rely on prediction techniques to estimate these attributes. Leveraging LLMs provides a robust approach to predicting these demographic attributes. By utilizing LLMs, we can leverage the extensive contextual understanding and knowledge encoded within the models to infer user demographics based on available information, such as personal descriptions and content within posts. The technical details are as follows.

\noindent \textbf{User Demographics Prediction with LLM.}
In order to predict user gender based on personal descriptions, since the collected data lacks sufficient labels, we use a public dataset released in~\cite{10.1145/3219819.3220077,10.1145/2700398} for assistance. It allows us to extract a vast array of labeled gender and personal description relationships. We filter out data with longer than 10 words in this dataset served as the ground truth to tune the language model. Specifically, we use ChatGLM~\cite{du-etal-2022-glm} as the foundation model and employ the P-Tuning-v2~\cite{liu-etal-2022-p} methodology. We feed the model with the personal description as a prompt and let the model determine the most probable gender associated with the given description.

To predict age using users' posts, we use Blog Authorship Corpus Dataset \cite{schler2006effects} dataset to establish the expression-to-age relationship. This dataset provides us with author-age labels for corresponding textual posts. We randomly select the historical blogs in \cite{schler2006effects} and add them to the prompt as input; then, the age can be used as the label for prefix tuning. The tuned large language model can be used to predict the age label in our collected social media dataset.

Next, we predict occupations only using pre-trained LLMs. In this scenario, we directly feed users' posts and personal profile descriptions to the LLM for prediction. By examining the content of these inputs, the model showcased its capacity to comprehend and infer users' occupations, further enhancing our demographic prediction capabilities.

\noindent \textbf{Prediction Result Evaluation}

The outcomes of our age and gender prediction analysis are presented in Table \ref{tab:dem_pre}. Our gender predictor, which relies on a fine-tuned Large Language Model (LLM), achieves satisfactory results. Despite the absence of explicit gender information in all personal descriptions, the predictor successfully generates valid predictions. Moving on to age, we select English blogs from \cite{schler2006effects} and ensured similar age distribution across the training and testing process. The results show that the mean squared error (MSE) was 128, while the mean absolute error (MAE) was around 7.53. These values indicate a 21.5\% unified percentage error (see Table \ref{tab:dem_pre}).

As for the occupations, we initially include the posts and personal descriptions of the combined user dataset in the prompt. We then feed the prompt to pre-trained ChatGLM to obtain the occupation of each user. We leave the supervised fine-tuning for occupation prediction as future work. It results in a total of 1,016 different occupations being identified from all users. However, utilizing all occupations is not essential since some occupations are very close. Thus, we group all occupations into 10 distinct occupation categories using the LLM, of which the categories can be found in Table \ref{tab:gpt_occ}. By condensing the number of occupations into a smaller set, we are able to simplify the simulation.

\begin{table}[t]
    \centering
    \begin{minipage}{0.45\textwidth}  %
            \caption{Prediction performance of gender and age.}
        \centering
        \small
        \renewcommand{\arraystretch}{1.2} %
        \begin{tabular}{c|ccc}
            \specialrule{1.5pt}{0pt}{0pt}
            \textbf{Demographic} & \multicolumn{3}{c}{\textbf{Performance}} \\ \midrule
            \multirow{2}{*}{Gender} & \textbf{Acc} & \textbf{F1} & \textbf{AUC} \\
            & 0.710 & 0.667 & 0.708 \\ \midrule
             \multirow{2}{*}{Age} & \textbf{MSE} & \textbf{MAE} & \textbf{Avg \% Error} \\
            & 128.0 & 7.53 & 21.50 \\ \specialrule{1.5pt}{0pt}{0pt}
        \end{tabular}
        \captionsetup{singlelinecheck=off, justification=centering}
        \label{tab:dem_pre}
    \end{minipage}
    \hfill
    \begin{minipage}{0.45\textwidth}  %
        \centering
        \caption{Ten occupations.}
        \renewcommand{\arraystretch}{1.2} %
        \begin{tabular}{|l|l|}
     \hline 1& \fontfamily{ppl}\selectfont Education Practitioner\\
 2&  \fontfamily{ppl}\selectfont Administrative Manager / Officer \\
3& \fontfamily{ppl}\selectfont Unemployed / Student \\
4& \fontfamily{ppl}\selectfont Engineer \\
5& \fontfamily{ppl}\selectfont Labor Technician / Worker \\
6& \fontfamily{ppl}\selectfont Logistics Practitioner \\
7& \fontfamily{ppl}\selectfont Medical Personnel \\
8& \fontfamily{ppl}\selectfont Financial Practitioner\\
9& \fontfamily{ppl}\selectfont Media Personnel \\
10& \fontfamily{ppl}\selectfont Entertainment and Arts Practitioner \\
\hline
        \end{tabular}
        \label{tab:gpt_occ}
    \end{minipage}
\end{table}

\subsection{Emotion and Attitude Simulation}

In our emotion simulation model, we adopt a Markov chain approach to capture the dynamic process of emotional changes triggered by a user receiving a message. The simulation involves four essential inputs: user demographics, current emotion, the received post. Emotions are classified into three distinct stages: calm, moderate, and intense.
User demographics serve as supplementary information LLMs, providing a reference point to contextualize emotional responses. The current emotion represents the user's emotional status before receiving the post, while the received post acts as the actuator for prompting the LLM to determine a new emotional status.

To regulate the decrease of emotional states over time, we introduce the decaying coefficient, a hyper-parameter that controls the decay rate of emotions. Our hypothesis assumes that emotions tend to diminish gradually as time passes, influencing the emotion simulation process. Throughout this intricate mechanism, we impart these details by prompt to the LLMs, which are responsible for deciding whether the emotional state should change in response to the received post. We are trying to reduce as much manual intervention as possible, to highlight the capability of LLMs in simulating emotional changes by posts. The attitude simulation is similar to the emotion simulation.

\subsection{Behavior Simulation}

\subsubsection{Content-generation Behavior}
In our social network simulation model, we incorporate an advanced approach utilizing Large Language Models (LLMs) to reproduce the dynamic process of content creation, shaped by users' emotions and attitudes towards specific events. The simulation hinges on two vital inputs: user profile information, and their current emotional or attitudinal state towards the event. Each piece of generated content is an embodiment of a user's internal state and external influences, reflecting their unique perspective.

User profile information serves as a reference point for the LLMs, furnishing essential context to shape content responses. The current emotional or attitudinal state symbolizes the user's mindset when reacting to the event, thereby playing a vital role in the LLM's generation of potential responses.

Underpinning this sophisticated mechanism is the profound cognitive and behavioral comprehension of LLMs. The LLM is prompted with these details and is then responsible for deciding how the content should be shaped in response to the event. Our aim is to minimize manual intervention as much as possible, to highlight the capability of LLMs in simulating authentic user-generated content.

The approach mirrors the way real-world users form their posts in response to distinct events, aligning the text generation process with the emotional or attitudinal dynamics of users. In this manner, we have been successful in utilizing LLMs to emulate the content creation process on social networks with high fidelity. 

\subsubsection{Interaction Behavior}

During the simulation, when a user receives a message from one of their followees, a critical decision needs to be made---whether to repost/post or not. 
That is to say, the interaction behavior includes reposting (forwarding) the original content and posting new content about the same social event.
The user's interaction behavior plays a pivotal role in propagating messages to the user's followers, facilitating the spread of information within the social network. However, modeling the complex mechanisms governing a user's interaction behavior poses significant challenges.
To address it, we employ large language models to capture the intricate relationship between the user, post features, and interaction behavior.

Specifically, to leverage the ability of LLMs to simulate a real user's interaction behavior, we prompt the model with information regarding the user's demographic properties, \textit{i.e.} gender, age, and occupation, in addition to the specific posts received, letting the LLM think like the user and make its decision. By such means, we enable LLM to make predictions regarding the user's inclination to repost the message or post new content.

To summarize, by employing the above approach, we can effectively harness the power of LLMs to predict users' interaction behavior, taking into account various user and post features.

\subsection{Other Implementation Details}
The system employs various techniques for utilizing or adapting large language models to the agent-based simulation.
For prompting-driven methods, we use either GPT-3.5 API provided by OpenAI\footnote{https://platform.openai.com/overview} or a ChatGLM-6B model~\cite{du-etal-2022-glm}.
For fine-tuning methods, we conduct the tuning based on the open-source ChatGLM model.

\section{Discussions and Open Problems}\label{sec::discussion}

The S$^3$ system, which has been developed, represents an initial endeavor aimed at harnessing the capabilities of large language models. This is to facilitate simulation within the domain of social science.

In light of this, our analysis delves further into its application and limitations, along with promising future improvements.

\subsection{Application of S$^3$ System}

Leveraging the powerful capabilities of large language models, this system excels in agent-based simulation. The system has the following applications in the field of social science.

\begin{itemize}[leftmargin=*]

    \item \textbf{Prediction.} Prediction is the most fundamental ability of agent-based simulation. Large language model-based simulation can be utilized to predict social phenomena, trends, and individual behaviors with historically collected data. For example, in economics, language models can help forecast market trends, predict consumer behavior, or estimate the impact of policy changes. In sociology, these models can aid in predicting social movements, public opinion shifts, or the adoption of new cultural practices. 

    \item \textbf{Reasoning and explanation.} During the simulation, each agent can be easily configured, and thus the system can facilitate reasoning and explanation in social science by generating phenomena with different configurations. Comparing the simulation results can provide explain the cause of the specific phenomena. Furthermore, the agent can be observed by prompts which can reflect how a human takes actions in the social environment.

    \item \textbf{Pattern discovery and theory construction.} With repeated simulation during the extremely less cost compared with real data collection, the simulation process can reveal some patterns of the social network. By uncovering patterns, these models can contribute to the development of new theories and insights. Furthermore, researchers can configure all the agents and the social network environment, based on an assumption or theory, and observe the simulation results. Testing the simulation results can help validate whether the proposed assumption or theory is correct or not.

    \item \textbf{Policy making.} The simulation can inform evidence-based policy-making by simulating and evaluating the potential outcomes of different policy interventions. It can assess the impact of policy changes on various social factors, including individual agents and the social environment. For example, in public health, it can simulate the spread of infectious diseases to evaluate the effectiveness of different intervention strategies. In urban planning, it can simulate the impact of transportation policies on traffic congestion or air pollution, by affecting how the users select public transportation. By generating simulations, these models can aid policymakers in making informed decisions.

\end{itemize}

\subsection{Improvement on Individual-level Simulation}
The current design of individual simulation still has several limitations requiring further improvement.
  First,  the agent requires more prior knowledge of user behavior, including how real humankind senses the social environment and makes decisions.  In other words, the simulation should encompass an understanding and integration of intricate contextual elements that exert influence on human behavior. Second, while prior knowledge of user behavior is essential, simulations also need to consider the broader context in which decisions are made. This includes factors such as historical events, social conditions, and personal experiences. By enhancing the agent's capacity to perceive and interpret contextual cues, more precise simulations can be achieved.

\subsection{Improvement on Population-level Simulation}

First, it is better to combine agent-based simulation with system dynamics-based methods.

Agent-based simulation focuses on modeling individual entities and their interactions, while system dynamics focuses on modeling the behavior of the social complex system as a whole. Through the fusion of these two methodologies, we can develop simulations of heightened comprehensiveness, encompassing both micro-level interactions and macro-level systemic behavior. This integration can provide a more accurate representation of population dynamics, including the impact of individual decisions on the overall system.

Second, we can consider a broader range of social phenomena. This involves modeling various societal, economic, and cultural factors that influence human behavior and interactions. Examples of social phenomena to consider include social networks, opinion dynamics, cultural diffusion, income inequality, and infectious disease spread. By incorporating these phenomena into the simulation, we can better validate the system's effectiveness and also gain more insights into social simulation.

\subsection{Improvement on System Architecture Design}

First, we can consider incorporating other channels for social event information. It is essential to acknowledge that social-connected users are not the sole providers of information for individuals within social networks. Consequently, the integration of supplementary data sources has the potential to enrich the individual simulation. For instance, recommender systems can be integrated to gather diverse information about social events. This integration can help capture a wider range of perspectives and increase the realism of the simulation.

Second, the system architecture should consider improving efficiency, which is essential for running large-scale simulations effectively. Optimizing the system architecture and computational processes can significantly enhance the performance and speed of simulations. To this end, techniques such as parallel computing, distributed computing, and algorithmic optimizations can be employed to reduce computational complexity and advance the efficiency of simulation runs. This allows for faster and more extensive exploration of scenarios, thereby enabling researchers to gain insights faster.

Third, it is essential to add an interface for policy intervention. Including an interface that allows policymakers to interact with the simulation can be beneficial. This interface would enable policymakers to input and test various interventions and policies in a controlled environment. By simulating the potential outcomes of different policy decisions, policymakers can make more informed choices. They can also evaluate the potential impact of their interventions on the simulated population. This feature can facilitate evidence-based decision-making and identify effective strategies.

\section{Conclusion}\label{sec::conclusion}
In this paper, we present the S$^3$ system (Social Network Simulation System) as a novel approach aimed at tackling the complexities of social network simulation. By harnessing the advanced capabilities of large language models (LLMs) in the realms of perception, cognition, and behavior, we have established a framework for social network emulation. Our simulations concentrate on three pivotal facets: emotion, attitude, and interactive behaviors. This research marks a significant stride forward in social network simulation, pioneering the integration of LLM-empowered agents. Beyond social science, our work possesses the potential to stimulate the development of simulation systems across diverse domains. Employing this methodology enables researchers and policymakers to attain profound insights into intricate social dynamics, thereby facilitating informed decision-making and effectively addressing various societal challenges.
\bibliographystyle{plain}
\bibliography{full}

\begin{thebibliography}{10}

\bibitem{aher2023using}
Gati~V Aher, Rosa~I Arriaga, and Adam~Tauman Kalai.
\newblock Using large language models to simulate multiple humans and replicate
  human subject studies.
\newblock In {\em International Conference on Machine Learning}, pages
  337--371. PMLR, 2023.

\bibitem{axelrod1997advancing}
Robert Axelrod.
\newblock Advancing the art of simulation in the social sciences.
\newblock In {\em Simulating social phenomena}, pages 21--40. Springer, 1997.

\bibitem{baumann2020modeling}
Fabian Baumann, Philipp Lorenz-Spreen, Igor~M Sokolov, and Michele Starnini.
\newblock Modeling echo chambers and polarization dynamics in social networks.
\newblock {\em Physical Review Letters}, 124(4):048301, 2020.

\bibitem{baumann2021emergence}
Fabian Baumann, Philipp Lorenz-Spreen, Igor~M Sokolov, and Michele Starnini.
\newblock Emergence of polarized ideological opinions in multidimensional topic
  spaces.
\newblock {\em Physical Review X}, 11(1):011012, 2021.

\bibitem{bratley1987guide}
Paul Bratley, Bennett~L Fox, and Linus~E Schrage.
\newblock A guide to simulation, 1987.

\bibitem{brown2020language}
Tom Brown, Benjamin Mann, Nick Ryder, Melanie Subbiah, Jared~D Kaplan, Prafulla
  Dhariwal, Arvind Neelakantan, Pranav Shyam, Girish Sastry, Amanda Askell,
  et~al.
\newblock Language models are few-shot learners.
\newblock {\em Advances in neural information processing systems},
  33:1877--1901, 2020.

\bibitem{charness2002understanding}
Gary Charness and Matthew Rabin.
\newblock Understanding social preferences with simple tests.
\newblock {\em The quarterly journal of economics}, 117(3):817--869, 2002.

\bibitem{chen2010scalable}
Wei Chen, Yifei Yuan, and Li~Zhang.
\newblock Scalable influence maximization in social networks under the linear
  threshold model.
\newblock In {\em 2010 IEEE international conference on data mining}, pages
  88--97. IEEE, 2010.

\bibitem{chopard1998cellular}
Bastien Chopard and Michel Droz.
\newblock Cellular automata.
\newblock {\em Modelling of Physical}, pages 6--13, 1998.

\bibitem{chowdhery2022palm}
Aakanksha Chowdhery, Sharan Narang, Jacob Devlin, Maarten Bosma, Gaurav Mishra,
  Adam Roberts, Paul Barham, Hyung~Won Chung, Charles Sutton, Sebastian
  Gehrmann, et~al.
\newblock Palm: Scaling language modeling with pathways.
\newblock {\em arXiv preprint arXiv:2204.02311}, 2022.

\bibitem{degroot1974reaching}
Morris~H DeGroot.
\newblock Reaching a consensus.
\newblock {\em Journal of the American Statistical association},
  69(345):118--121, 1974.

\bibitem{du-etal-2022-glm}
Zhengxiao Du, Yujie Qian, Xiao Liu, Ming Ding, Jiezhong Qiu, Zhilin Yang, and
  Jie Tang.
\newblock {GLM}: General language model pretraining with autoregressive blank
  infilling.
\newblock In {\em Proceedings of the 60th Annual Meeting of the Association for
  Computational Linguistics (Volume 1: Long Papers)}, pages 320--335, Dublin,
  Ireland, May 2022. Association for Computational Linguistics.

\bibitem{anil2023palm}
Rohan~Anil et~al.
\newblock Palm 2 technical report, 2023.

\bibitem{flamino2023political}
James Flamino, Alessandro Galeazzi, Stuart Feldman, Michael~W Macy, Brendan
  Cross, Zhenkun Zhou, Matteo Serafino, Alexandre Bovet, Hern{\'a}n~A Makse,
  and Boleslaw~K Szymanski.
\newblock Political polarization of news media and influencers on twitter in
  the 2016 and 2020 us presidential elections.
\newblock {\em Nature Human Behaviour}, pages 1--13, 2023.

\bibitem{forrester1993system}
Jay~W Forrester.
\newblock System dynamics and the lessons of 35 years.
\newblock In {\em A systems-based approach to policymaking}, pages 199--240.
  Springer, 1993.

\bibitem{gilbert2005simulation}
Nigel Gilbert and Klaus Troitzsch.
\newblock {\em Simulation for the social scientist}.
\newblock McGraw-Hill Education (UK), 2005.

\bibitem{hamalainen2023evaluating}
Perttu H{\"a}m{\"a}l{\"a}inen, Mikke Tavast, and Anton Kunnari.
\newblock Evaluating large language models in generating synthetic hci research
  data: a case study.
\newblock In {\em Proceedings of the 2023 CHI Conference on Human Factors in
  Computing Systems}, pages 1--19, 2023.

\bibitem{he2014minimum}
Jing He, Shouling Ji, Raheem Beyah, and Zhipeng Cai.
\newblock Minimum-sized influential node set selection for social networks
  under the independent cascade model.
\newblock In {\em Proceedings of the 15th ACM International Symposium on Mobile
  ad hoc Networking and Computing}, pages 93--102, 2014.

\bibitem{hohmann2023quantifying}
Marilena Hohmann, Karel Devriendt, and Michele Coscia.
\newblock Quantifying ideological polarization on a network using generalized
  euclidean distance.
\newblock {\em Science Advances}, 9(9):eabq2044, 2023.

\bibitem{horton2023large}
John~J Horton.
\newblock Large language models as simulated economic agents: What can we learn
  from homo silicus?
\newblock Technical report, National Bureau of Economic Research, 2023.

\bibitem{kolesar1975simulation}
Peter Kolesar and Warren~E Walker.
\newblock A simulation model of police patrol operations: program description.
\newblock 1975.

\bibitem{lee2021all}
Lik-Hang Lee, Tristan Braud, Pengyuan Zhou, Lin Wang, Dianlei Xu, Zijun Lin,
  Abhishek Kumar, Carlos Bermejo, and Pan Hui.
\newblock All one needs to know about metaverse: A complete survey on
  technological singularity, virtual ecosystem, and research agenda.
\newblock {\em arXiv preprint arXiv:2110.05352}, 2021.

\bibitem{liu2023emergence}
Jiazhen Liu, Shengda Huang, Nathaniel~M Aden, Neil~F Johnson, and Chaoming
  Song.
\newblock Emergence of polarization in coevolving networks.
\newblock {\em Physical Review Letters}, 130(3):037401, 2023.

\bibitem{liu-etal-2022-p}
Xiao Liu, Kaixuan Ji, Yicheng Fu, Weng Tam, Zhengxiao Du, Zhilin Yang, and Jie
  Tang.
\newblock {P}-tuning: Prompt tuning can be comparable to fine-tuning across
  scales and tasks.
\newblock In {\em Proceedings of the 60th Annual Meeting of the Association for
  Computational Linguistics (Volume 2: Short Papers)}, pages 61--68, Dublin,
  Ireland, May 2022. Association for Computational Linguistics.

\bibitem{lorenz2023systematic}
Philipp Lorenz-Spreen, Lisa Oswald, Stephan Lewandowsky, and Ralph Hertwig.
\newblock A systematic review of worldwide causal and correlational evidence on
  digital media and democracy.
\newblock {\em Nature human behaviour}, 7(1):74--101, 2023.

\bibitem{luding2005information}
Stefan Luding.
\newblock Information propagation.
\newblock {\em Nature}, 435(7039):159--160, 2005.

\bibitem{marsh1978using}
Lawrence~C Marsh and Meredith Scovill.
\newblock Using system dynamics to model the social security system.
\newblock In {\em NBER Workshop on Policy Analysis with Social Security
  Research Files}, pages 15--17, 1978.

\bibitem{meadows1974dynamics}
Dennis~L Meadows, William~W Behrens, Donella~H Meadows, Roger~F Naill,
  J{\o}rgen Randers, and Erich Zahn.
\newblock {\em Dynamics of growth in a finite world}.
\newblock Wright-Allen Press Cambridge, MA, 1974.

\bibitem{notarmuzi2022universality}
Daniele Notarmuzi, Claudio Castellano, Alessandro Flammini, Dario Mazzilli, and
  Filippo Radicchi.
\newblock Universality, criticality and complexity of information propagation
  in social media.
\newblock {\em Nature communications}, 13(1):1308, 2022.

\bibitem{okawa2022predicting}
Maya Okawa and Tomoharu Iwata.
\newblock Predicting opinion dynamics via sociologically-informed neural
  networks.
\newblock In {\em Proceedings of the 28th ACM SIGKDD conference on knowledge
  discovery and data mining}, pages 1306--1316, 2022.

\bibitem{openai2023gpt4}
OpenAI.
\newblock Gpt-4 technical report, 2023.

\bibitem{park2023generative}
Joon~Sung Park, Joseph~C O'Brien, Carrie~J Cai, Meredith~Ringel Morris, Percy
  Liang, and Michael~S Bernstein.
\newblock Generative agents: Interactive simulacra of human behavior.
\newblock {\em arXiv preprint arXiv:2304.03442}, 2023.

\bibitem{10.1145/3219819.3220077}
Jiezhong Qiu, Jian Tang, Hao Ma, Yuxiao Dong, Kuansan Wang, and Jie Tang.
\newblock Deepinf: Social influence prediction with deep learning.
\newblock In {\em Proceedings of the 24th ACM SIGKDD International Conference
  on Knowledge Discovery and Data Mining}, KDD '18, page 2110–2119, New York,
  NY, USA, 2018. Association for Computing Machinery.

\bibitem{samuelson1988status}
William Samuelson and Richard Zeckhauser.
\newblock Status quo bias in decision making.
\newblock {\em Journal of risk and uncertainty}, 1:7--59, 1988.

\bibitem{santos2021link}
Fernando~P Santos, Yphtach Lelkes, and Simon~A Levin.
\newblock Link recommendation algorithms and dynamics of polarization in online
  social networks.
\newblock {\em Proceedings of the National Academy of Sciences},
  118(50):e2102141118, 2021.

\bibitem{schafer2002spinning}
Joseph~A Schafer.
\newblock Spinning the web of hate: Web-based hate propagation by extremist
  organizations.
\newblock {\em Journal of Criminal Justice and Popular Culture}, 2002.

\bibitem{schler2006effects}
Jonathan Schler, Moshe Koppel, Shlomo Argamon, and James~W Pennebaker.
\newblock Effects of age and gender on blogging.
\newblock In {\em AAAI spring symposium: Computational approaches to analyzing
  weblogs}, volume~6, pages 199--205, 2006.

\bibitem{spencer1984effect}
Peter~D Spencer.
\newblock The effect of oil discoveries on the british economy—theoretical
  ambiguities and the consistent expectations simulation approach.
\newblock {\em The Economic Journal}, 94(375):633--644, 1984.

\bibitem{touvron2023llama}
Hugo Touvron, Thibaut Lavril, Gautier Izacard, Xavier Martinet, Marie-Anne
  Lachaux, Timoth{\'e}e Lacroix, Baptiste Rozi{\`e}re, Naman Goyal, Eric
  Hambro, Faisal Azhar, et~al.
\newblock Llama: Open and efficient foundation language models.
\newblock {\em arXiv preprint arXiv:2302.13971}, 2023.

\bibitem{troitzsch1996social}
Klaus~G Troitzsch.
\newblock Social science microsimulation.
\newblock Springer Science \& Business Media, 1996.

\bibitem{wang2022global}
Jianghao Wang, Yichun Fan, Juan Palacios, Yuchen Chai, Nicolas
  Guetta-Jeanrenaud, Nick Obradovich, Chenghu Zhou, and Siqi Zheng.
\newblock Global evidence of expressed sentiment alterations during the
  covid-19 pandemic.
\newblock {\em Nature Human Behaviour}, 6(3):349--358, 2022.

\bibitem{xie2021detecting}
Jiarong Xie, Fanhui Meng, Jiachen Sun, Xiao Ma, Gang Yan, and Yanqing Hu.
\newblock Detecting and modelling real percolation and phase transitions of
  information on social media.
\newblock {\em Nature Human Behaviour}, 5(9):1161--1168, 2021.

\bibitem{yildiz2010voting}
Mehmet~E Yildiz, Roberto Pagliari, Asuman Ozdaglar, and Anna Scaglione.
\newblock Voting models in random networks.
\newblock In {\em 2010 information theory and applications workshop (ITA)},
  pages 1--7. IEEE, 2010.

\bibitem{zang2020neural}
Chengxi Zang and Fei Wang.
\newblock Neural dynamics on complex networks.
\newblock In {\em Proceedings of the 26th ACM SIGKDD international conference
  on knowledge discovery \& data mining}, pages 892--902, 2020.

\bibitem{zeng2022glm}
Aohan Zeng, Xiao Liu, Zhengxiao Du, Zihan Wang, Hanyu Lai, Ming Ding, Zhuoyi
  Yang, Yifan Xu, Wendi Zheng, Xiao Xia, et~al.
\newblock Glm-130b: An open bilingual pre-trained model.
\newblock {\em arXiv preprint arXiv:2210.02414}, 2022.

\bibitem{10.1145/2700398}
Jing Zhang, Jie Tang, Juanzi Li, Yang Liu, and Chunxiao Xing.
\newblock Who influenced you? predicting retweet via social influence locality.
\newblock {\em ACM Trans. Knowl. Discov. Data}, 9(3), apr 2015.

\end{thebibliography}

\end{document}